\documentclass[prd,preprint,nofootinbib]{revtex4}

\usepackage{graphicx}
\usepackage{amssymb}
\usepackage{amsmath}

\preprint{IPMU-10-0160}
\preprint{KEK-TH-1401}

\begin{document}

\newcommand{\beq}{\begin{equation}}   
\newcommand{\eeq}{\end{equation}}
\newcommand{\bea}{\begin{eqnarray}}   
\newcommand{\eea}{\end{eqnarray}}
\newcommand{\bear}{\begin{array}}  
\newcommand {\eear}{\end{array}}
\newcommand{\bef}{\begin{figure}}  
\newcommand {\eef}{\end{figure}}
\newcommand{\bec}{\begin{center}}  
\newcommand {\eec}{\end{center}}
\newcommand{\non}{\nonumber}  
\newcommand {\eqn}[1]{\beq {#1}\eeq}
\newcommand{\la}{\left\langle}  
\newcommand{\ra}{\right\rangle}
\newcommand{\ds}{\displaystyle}
\def\SEC#1{Sec.~\ref{#1}}
\def\FIG#1{Fig.~\ref{#1}}
\def\EQ#1{Eq.~(\ref{#1})}
\def\EQS#1{Eqs.~(\ref{#1})}
\def\GEV#1{10^{#1}{\rm\,GeV}}
\def\MEV#1{10^{#1}{\rm\,MeV}}
\def\KEV#1{10^{#1}{\rm\,keV}}
\def\lrf#1#2{ \left(\frac{#1}{#2}\right)}
\def\lrfp#1#2#3{ \left(\frac{#1}{#2} \right)^{#3}}
\newcommand{\phih}{\hat{\phi}}
\newcommand{\phit}{\tilde{\phi}}
\newcommand{\phid}{\phi^{\dag}}
\newcommand{\phib}{\bar{\phi}}
\newcommand{\vp}{\varphi}

%

\title{
General Analysis of Inflation in the Jordan frame Supergravity
}

\author{
Kazunori Nakayama${}^{(a)}$ and
Fuminobu Takahashi${}^{(b)}$
}

\affiliation{
${}^{(a)}$ Theory Center, KEK, 1-1 Oho, Tsukuba, Ibaraki 305-0801, Japan\\
${}^{(b)}$ Institute for the Physics and Mathematics of the Universe,
University of Tokyo, Chiba 277-8583, Japan
}

\date{\today}

\begin{abstract}
  We study various inflation models in the Jordan frame supergravity
  with a logarithmic K\"ahler potential. We find that, in a class of
  inflation models containing an additional singlet in the
  superpotential, three types of inflation can be realized: the
  Higgs-type inflation, power-law inflation, and chaotic inflation
  with/without a running kinetic term. The former two are possible if
  the holomorphic function dominates over the non-holomorphic one in
  the frame function, while the chaotic inflation occurs when both are
  comparable. Interestingly, the fractional-power potential can be
  realized by the running kinetic term.  We also discuss the
  implication for the Higgs inflation in supergravity.
\end{abstract}

\pacs{98.80.Cq}

\maketitle

\section{Introduction}
The inflation is strongly motivated by the recent WMAP
results~\cite{Komatsu:2010fb}.  However, it is a non-trivial task to
construct a successful inflation model, partly because the properties
of the inflaton are poorly known.

Recently, a new class of inflation models was proposed by one of the
authors (FT)~\cite{Takahashi:2010ky}, in which the kinetic term grows
as the inflaton field, making the effective potential
flat~\cite{Dimopoulos:2003iy,Izawa:2007qa}.  This model naturally fits
with a high-scale inflation model such as chaotic
inflation~\cite{Linde:1983gd}, in which the inflaton moves over a
Planck scale or even larger within the last $50$ or $60$
e-foldings~\cite{Lyth:1996im}.  This is because the precise form of
the kinetic term may well change after the inflaton travels such a
long distance.  In some cases, the change could be so rapid, that it
significantly affects the inflaton dynamics.  We named such model as
running kinetic (RK) inflation.  In order to realize a chaotic
inflation in supergravity, some sort of shift symmetry is necessary.
One way to implement the RK inflation model in supergravity is to
impose a shift symmetry on a composite field:\footnote{ The shift
  symmetry is sufficient but not necessary for having the RK
  inflation, and a more general form of the K\"ahler potential leads
  to the RK inflaiton.  }
\bea
\phi^n \;\rightarrow\;\phi^n+ \alpha,
\label{sym}
\eea
where $\alpha \in {\bf R}$ is a transformation parameter, $n$ is a
positive integer, and we adopt the Planck unit in which $M_P = 2.4
\times \GEV{18}$ is set to be unity.  If $n=1$, this symmetry is
reduced to that considered in Ref.~\cite{Kawasaki:2000yn}.
Interestingly, the power of the inflaton potential generically changes
in the RK inflation models, which makes it possible to realize chaotic
inflation with e.g. a linear and fractional-power
potential~\cite{Takahashi:2010ky}.  The phenomenological aspects of
the RK inflation was studied in detail in Ref.~\cite{Nakayama:2010kt},
and the idea led to a new Higgs chaotic inflation in
supergravity~\cite{Nakayama:2010sk}.

Another way to obtain a flat potential is to introduce a non-minimal
coupling to
gravity~\cite{Salopek:1988qh,Futamase:1987ua,Spokoiny:1984bd,Fakir:1990eg,Komatsu:1997hv}.
This idea has recently attracted much attention since the proposal of
the standard model (SM) Higgs inflation~\cite{Bezrukov:2007ep}.  There
are studies on the Higgs inflation in supergravity with the same
spirit~\cite{Einhorn:2009bh,Lee:2010hj,Ferrara:2010yw,Ferrara:2010in,Kallosh:2010ug}. In
the Jordan frame supergravity, the non-minimal coupling to gravity is
represented by a holomorphic function $J(z)$ and a generic
non-holomorphic function $g(z,{\bar z})$ in the frame function
$\Omega^2(z,{\bar z})$:
\bea
\frac{1}{\sqrt{-g}}{\cal L}_{\rm grav} &=&\frac{1}{2}\Omega^2(z,{\bar z}) R + \cdots,\\
\Omega^2(z,{\bar z}) &=& 1 - \frac{1}{3} \left(
g(z,{\bar z})
+ J(z) 
+ \bar{J}({\bar z})\right), 
\eea
where $R$ denotes a curvature scalar, $z$ and ${\bar z}$ are complex
scalar fields, and $g(z,{\bar z})$ and $J(z)$ are non-holomorphic and
holomorphic functions, respectively.  If $g(z,{\bar z}) = |z|^2$ and
$J(z) = 0$, $z$ has a canonical kinetic term with a conformal coupling
to gravity.  The frame function is related to the K\"ahler potential
as
\beq
K(z,\bar{z}) = -3 \log \Omega^2(z,\bar{z}).
\label{kahler-omega}
\eeq
In Ref.~\cite{Kallosh:2010ug}, they studied various inflation models and one of them is such that
$g(z,{\bar z}) = |z|^2$ and $J(z) = \frac{3 \chi}{4} z^2$ with $\chi =
\pm 2/3$, which exhibits an (accidental) shift symmetry on
$z$~\footnote{ In Ref.~\cite{BenDayan:2010yz}, a shift symmetry on
  $H_u$ and $H_d$ is introduced to fortify the form of the K\"ahler
  potential. However, the resultant potential is a quartic power of
  the inflaton, which is severely constrained by the WMAP
  observation~\cite{Komatsu:2010fb}. In fact, the model of
  Ref.~\cite{BenDayan:2010yz} is similar to the early work on the
  inflation using the MSSM flat direction~\cite{Kasuya:2003iv}.  }.

One of the purposes of this letter is to investigate what kind of
inflation models are possible in the Jordan frame supergravity with a
logarithmic K\"ahler potential. In particular, we would like to
clarify the relation among the RK inflation, inflation with
non-minimal coupling to gravity, and the chaotic inflation with an
accidental shift symmetry.  Also, the analysis on the RK inflation was
performed with a polynomial K\"ahler potential so far, and it is a
non-trivial question whether the RK inflation occurs with a
logarithmic K\"ahler potential.

In this letter we study inflation in the Jordan frame supergravity
with a logarithmic K\"ahler potential, which contains general
holomorphic and non-holomorphic functions. For superpotential, we
introduce an additional singlet $X$ to have a successful chaotic
inflaiton~\cite{Kawasaki:2000yn}.  Focusing on a large-field inflation
model, we find that three types of inflation are possible in this
framework, namely, the Higgs-type inflation, the power-law
inflation~\cite{Abbott:1984fp}, and the chaotic inflation with/without
a running kinetic term.  It is interesting that all the three
inflation models can be consistent with the current CMB observations.
In the next section we will study the inflation models with a
logarithmic K\"ahler potential, and Sec.~3 is devoted to discussion
and conclusions.

\section{Analysis}

Let us  consider the frame function in the following form,
\beq
\Omega^2\;=\; 1 - \frac{1}{3} \left( g(\phi,{\bar \phi})+ |X|^2 + \zeta |X|^4
+ J(\phi) 
+ \bar{J}({\bar \phi})\right),
\label{Omega2}
\eeq
where $\phi$ is the inflaton, and $X$ is a singlet field. 
The superpotential is given by~\footnote{This form is a natural extension of the interaction
proposed in Ref.~\cite{Kawasaki:2000yn}. The superpotential (\ref{Xphi})
was considered in Ref.~\cite{Takahashi:2010ky}, and recently it was also studied in 
Refs.~\cite{Nakayama:2010kt,Kallosh:2010ug,Nakayama:2010sk}.
}
\beq
W\;=\; \lambda X \phi^m,
\label{Xphi}
\eeq
where $\lambda$ is a coupling constant, and $m$ is a positive integer. Using the phase degree of
freedom of $X$, we take $\lambda$ to be real and positive. 
The presence of $X$ is essential for constructing chaotic inflation in  supergravity,
and it makes the form of the scalar potential simple. Since the $X$ can be stabilized
at the origin by the quartic coupling in $\Omega$~\cite{Lee:2010hj}, 
we will set $\la X \ra = 0$ in the following analysis. 

We consider the following three cases: 
\begin{center}
\begin{enumerate}
\item $g(\phi,{\bar \phi}) \gg |J(\phi)|$,
\item $g(\phi,{\bar \phi}) \ll |J(\phi)|$,
\item $g(\phi,{\bar \phi}) \sim |J(\phi)|$,
\end{enumerate}
\end{center}
where the inequalities are estimated during inflation.  To simplify
the analysis, we focus on the case that $g$ and $J$ can be
approximated as a power of $\phi$ and ${\bar \phi}$ during the
relevant epoch of the inflation:
\bea
g(\phi,{\bar \phi}) &\approx& |\phi|^2+a |\phi|^{2 \ell},\\
J(\phi) &\approx&b\, \phi^n,
\eea
where $a$ and $b$ are real and complex parameters, respectively, and
$\ell>1$ and $n$ are positive integers. 
Here we allow the non-holomorphic function $g(\phi, \bar \phi)$ to take a general form,
because the kinetic term could change during inflation especially if the inflaton
travels a large distance. Note that the inflation is still canonically normalized about the origin.
We will set $b$ to be real and
positive by re-defining the phase of $\phi$ without loss of
generality.  In a more general case, there could be other scheme that
is not described by our analysis.  As a first step, however, the above
three cases cover reasonably large portion of the possible inflation
models.
 
Before going to the analysis on each case, we here show the K\"ahler
metric and the scalar potential for the inflaton:
\bea
{\cal L}&=& K_{\phi {\bar \phi}} \partial \phi^\dag \partial \phi - V(\phi,\phi^\dag),
\eea
with
\bea
\label{K}
K_{\phi {\bar \phi}} &=&\frac{1}{\Omega^4}\left(
1-\frac{1}{3}a (\ell-1)^2 |\phi|^{2\ell}+a \ell^2|\phi|^{2\ell-2}+\frac{1}{3}b(n-1)(\phi^n+\phi^{\dag n})
\right.\non\\
&&~~~~~~\left.
+\,\frac{1}{3}b^2n^2|\phi|^{2n-2}
-\frac{1}{3} a b \ell (\ell-n) |\phi|^{2\ell-2}(\phi^n+\phi^{\dag n})
\right),\\
\label{V}
V(\phi,\phi^\dag)&=& \frac{ \lambda^2 |\phi|^{2m}}{\Omega^4},\\
\Omega^2&=&1 - \frac{1}{3} \left( |\phi|^2 +a |\phi|^{2 \ell} + b \phi^n + b \phi^{\dag n}\right),
\eea
where we set $\la X \ra = 0$. Here and in what follows we adopt the Einstein frame.
 
\subsection{A case of $g(\phi,{\bar \phi}) \gg |J(\phi)|$}
First let us consider the case of $g(\phi,{\bar \phi}) \gg |J(\phi)|$.
In this limit, we can set $b=0$. As we will see below, the inflation
does not take place in this case. The K\"ahler metric and the frame
function are given by
\bea
K_{\phi {\bar \phi}} &=&\frac{1}{\Omega^4}\left(
1-\frac{1}{3}a (\ell-1)^2 |\phi|^{2\ell}+a \ell^2|\phi|^{2\ell-2}\right),\\
\Omega^2&=&1 - \frac{1}{3} \left( |\phi|^2 +a |\phi|^{2 \ell}\right).
\eea
For $a>0$, it is clear from the expression of $\Omega^2$, $\phi$
cannot take a value much larger than $O(1)$, since otherwise
$\Omega^2$ becomes negative and unphysical. We can easily see that
both the K\"ahler metric and the potential $V(\phi,\phi^\dag)$ diverge
where $\Omega^2$ vanishes.  (The numerator of the K\"ahler metric does
not vanish at this point).  Let us estimate the effective potential in
terms of a canonically normalized field near the point where $\Omega^2
= 0$, in case of $a=0$. The situation is similar (actually even worse)
in the case of $a>0$. Let us define $\varphi \equiv |\phi|$.  The
canonically normalized field $\hat{\varphi}$ is related to $\varphi$
as
\bea
\hat{\varphi}&=&\int \sqrt{2 K_{\phi {\bar \phi}}}~ d \varphi \\
                     &\approx& \sqrt{\frac{3}{2}} \log\lrf{2\sqrt{3}}{\sqrt{3}-\varphi} ~~~{\rm as}~~~\varphi \rightarrow \sqrt{3}.
\eea
As $\varphi$ approaches $\sqrt{3}$, the canonically normalized field
${\hat \varphi}$ goes to infinity.  At a sufficiently large ${\hat
  \varphi}$, the potential is approximated with~\cite{Futamase:1987ua}
\beq
V({\hat \varphi})\;\simeq\;\frac{3^m \lambda^2}{16} \exp\left({\sqrt{\frac{8}{3}} {\hat \varphi}}\right).
\eeq
Thus, the effective potential is an exponentially growing function and
the inflation does not occur.

If $a$ is negative and sufficiently large, $\Omega^2$ does not vanish
at a large value of $\varphi$.  However, in this case, the K\"ahler
metric necessarily vanishes at a finite value of $\varphi \gtrsim
O(1)$, and the inflaton will be strongly coupled near the point.  The
inflation does not occur in this case, either.

The fact that the inflation does not occur in this case strongly
motivates us to introduce a holomorphic function $J(\phi)$, which
should play an important role for the inflation.  Interestingly, two
different types of inflation are possible depending on the relative
size of the holomorphic and non-holomorphic functions.

\subsection{A case of $g(\phi,{\bar \phi}) \ll |J(\phi)|$}   \label{case2}
Secondly we consider a case that the holomorphic function $J(\phi)$
dominates over the non-holomorphic function $g(\phi,{\bar
  \phi})$. This is the case that the non-minimal coupling to gravity
plays an important role, and the Higgs inflation in the
next-to-minimal supersymmetric standard model (NMSSM) falls in this
category.  For simplicity we set $a=0$, and the situation is
qualitatively similar in the case of $a \ne 0$.  The K\"ahler metric
and the frame function are given by
\bea
\label{case2:K}
K_{\phi {\bar \phi}} &=&\frac{1}{\Omega^4}\left(
1
+\frac{1}{3}b(n-1)(\phi^n+\phi^{\dag n})
+\,\frac{1}{3}b^2n^2|\phi|^{2n-2}
\right),\\
\Omega^2&=&1 - \frac{1}{3} \left( |\phi|^2 + b \phi^n + b \phi^{\dag n}\right).
\label{Omega_case2}
\eea
Note that, in contrast to the previous case, there are directions in
the field space of $\phi$ such that both $\Omega^2$ and K\"ahler
metric neither vanish nor diverge at finite values of $\phi$.  Since
the $\Omega^2$ appears in the denominator of the potential $V$ in
Eq.~(\ref{V}), the phase of $\phi$ is stabilized so that
$\phi^n+\phi^{\dag n}$ takes the minimal value, for sufficiently large
$\phi$. (Remember that we set $b>0$). To see this let us decompose
$\phi = \vp e^{i \theta}$. Then the frame function is given by
\beq
\Omega^2\;=\;1-\frac{1}{3} \vp^2-\frac{2b}{3} \vp^n \cos n\theta,
\eeq
and therefore the potential is minimized at~\footnote{
The mass of the phase is of $O(H)$ for $n = O(1)$. However, it becomes light for $n \gtrsim O(10)$,
which may result in the isocurvature perturbation or non-Gaussianity.
}
\beq
\theta_{\rm min} \;=\; \frac{\pi}{n}(2k+1)
\label{theta_min}
\eeq
with $k=0,\cdots n-1$.  Along the inflation trajectory given by
(\ref{theta_min}), the radial component $\varphi$ can take a
super-Planckian value. This is because, when the holomorphic function
is large enough, there is an approximate shift symmetry on $J(\phi) =
b \phi^n$ in the frame function (\ref{Omega_case2}), namely,
\beq
 \phi^n \rightarrow \phi^n + i \alpha,
\eeq
which is equivalent to (\ref{sym}). It is remarkable that a shift
symmetry on a composite field $\phi^n$ appears in the limit that the
holomorphic function becomes large, namely, the non-minimal coupling
to gravity gets large. Indeed, the inflationary trajectory
(\ref{theta_min}) coincides with that of the RK inflation considered
in Ref.~\cite{Nakayama:2010kt}. However, the form of the kinetic term
is not same because of the logarithmic K\"ahler potential. (If the
last term in the numerator in Eq.~(\ref{case2:K}) dominated and if
$\Omega^2$ were a constant, the kinetic term would grow at large
$\phi$, leading to the RK inflation.)  As we will see shortly, the RK
inflation is realized when the holomorphic function is comparable to
the non-holomorphic one.

Let us comment on the lower bound on $b$. For $\Omega^2$ not to vanish along the trajectory (\ref{theta_min}),
$b$ must be larger than $b_c$ given by
\beq
b\;>\; b_c \equiv \left\{
\bear{cc}
\ds{\frac{(n-2)^\frac{n-2}{2}}{3^\frac{n-2}{2} n^{\frac{n}{2}}}}&{\rm~~~for~~}n>2\\
&\\
\ds{\frac{1}{2}} & {\rm~~~for~~}n=2
\eear
\right.
\label{ineq1}
\eeq
If this inequality is met, the K\"ahler metric does not diverge at a finite value of $\varphi$.
Furthermore, in order for the scalar potential not to have a local maximum (and minimum),
$b$ must be larger than $b_c^\prime$,
\beq
b\;>\;b_c^\prime \equiv \frac{(m-2)^\frac{n}{2}}{m^\frac{n-2}{2} (m-n)} b_c,
\label{ineq2}
\eeq
where we have assumed $m > n$. For $m > n \geq 2$, $b_c^\prime$ is greater than or equal to $b_c$.
As we will see below, if $m < n$, the potential has a local maximum and exhibits runaway behavior.
If $m=n \geq 3$, there is always local maximum for any $b>b_c$.
If $m=n=2$, there is no local maximum for any $b>b_c$.
We assume (\ref{ineq1}) is satisfied in the following.

Using (\ref{theta_min}), we obtain the Lagrangian,
\beq
{\cal L}\;=\; \ds{ \frac{1-\frac{2b}{3}(n-1) \varphi^n+\frac{b^2}{3}n^2 \varphi^{2n-2}}{\left(
1-\frac{1}{3}\varphi^2+\frac{2b}{3} \varphi^n\right)^2}} (\partial \varphi)^2
-  \frac{ \lambda^2 \varphi^{2m}}{\left(1-\frac{1}{3}\varphi^2+\frac{2b}{3} \varphi^n\right)^2}
\label{Lagrang}
\eeq
Let us consider the limit $\varphi \gg b^{-1/(n-2)}$, in which case the above Lagrangian 
is simplified as
\beq
{\cal L}\; \approx \; \frac{3n^2}{4 } \varphi^{-2}\partial \varphi^2
-  \frac{9 \lambda^2}{4b^2} \varphi^{2(m-n)} \left(1-\frac{1}{2b} \varphi^{2-n}+\frac{3}{2b} \varphi^{-n}\right)^{-2}.
\eeq
The canonically normalized inflaton ${\hat \varphi}$ is related to $\varphi$ as~\footnote{Precisely speaking,
we need to introduce a scale $M$ in the logarithmic function, which results in a shift of $\varphi$. This does
not affect the following discussion.}
\beq
{\hat \varphi}\;\approx\; \sqrt{\frac{3}{2}}n \ln(\varphi),
\eeq
and the scalar potential in terms of ${\hat \varphi}$ is given by
\beq
V({\hat \varphi}) \approx \frac{9 \lambda^2}{4b^2} e^{2\alpha \frac{m-n}{n} \hat{\varphi}}
 \left(1-\frac{1}{2b} e^{\alpha \frac{2-n}{n}{\hat \varphi}}+\frac{3}{2b} e^{-\alpha {\hat \varphi}}\right)^{-2},
\eeq
where we defined $\alpha \equiv \sqrt{2/3}$. 
As we have mentioned, the potential $V$ exhibits  runaway behavior for $m < n$ as well as $m=n \geq 3$,
while the potential is an exponentially  growing function for $m > n$. 
If $m=n=2$, the scalar potential asymptotically approaches
a constant value and the tilt of the potential is exponentially suppressed. The last case corresponds to the Higgs inflation~\cite{Bezrukov:2007ep,
Einhorn:2009bh,Ferrara:2010yw,Lee:2010hj,Ferrara:2010in,Kallosh:2010ug}, and it was extensively studied in the literatures, and so, we do not repeat the analysis here.

Let us consider the case of $m>n$. In this case the scalar potential
grows exponentially, and so, one might think that no inflation occurs
in this case. However, the inflation does occur if the coefficient in
the exponent is small enough. This is actually the power-law inflation
with a positive exponent.  The slow-roll parameters $\varepsilon$ and
$\eta$ are given by
\bea
\varepsilon &= & \frac{4}{3} \lrfp{m-n}{n}{2},~~~\eta \;=\;   \frac{8}{3} \lrfp{m-n}{n}{2}.
\eea
Therefore the inflation occurs if $(m-n)/n \ll 1$. Note that this is
not a severe tuning of parameters; $(m-n)/n \sim 0.1$ is
sufficient. Such a choice of $m$ and $n$ can be justified for a
certain choice of discrete and U(1)$_R$ symmetries.  The inflaton
field is related to the e-folding number $N$ as
\beq
{\hat \varphi} \;\simeq\;  \sqrt{\frac{8}{3}} \frac{m-n}{n} N
\eeq
Interestingly, the tensor-to-scalar ratio $r$ and the scalar spectral
index do not depend on the e-folding number, and they are determined
by $m$ and $n$:
\beq
n_s \;\simeq\; 1- \frac{8}{3} \lrfp{m-n}{n}{2},~~~r\;\simeq\;\frac{64}{3} \lrfp{m-n}{n}{2},
\eeq
and
\beq
1-n_s \;=\; \frac{r}{8}.
\eeq
We note that the power-law inflation ends when $\varphi \sim
b^{-1/(n-2)}$. If $1>b>b_c^\prime$, a chaotic inflation with a
potential $\propto \varphi^{2m}$ occurs after the power-law inflation.
Since $n_s$ and $r$ do not depend on the duration of the power-law
inflation, the above prediction is not changed in this case, if the e-folds of the chaotic inflation is smaller than $50$.

Thus, the inflation model considered here can be either the power-law
inflation ($0<(m-n)/n \ll 1$) or the Higgs-type inflation ($m=n=2$).

\subsection{A case of $g(\phi,{\bar \phi}) \sim |J(\phi)|$}

Thirdly we consider a case of $g(\phi,{\bar \phi}) \sim
|J(\phi)|$. For the equality to hold for a reasonably large field
space, we take 
(i) $a=0$ and $b=1/2$ for $n=2$, or
(ii) $|a| \sim b$ and $2 \ell = n$ for $n>2$. 
As we shall see, depending on the value of $a$, there appears an approximate flat
direction corresponding to a shift symmetry on $\phi^\ell$, which
results in the RK inflation.

First consider the case (i) ($n=2$). 
If we take $a=0$ and $b\to1/2$, we find that there appears an accidental shift symmetry $\phi \to \phi + i\alpha$,
noting that
\bea
	K_{\phi\bar \phi} &=& \frac{1}{\Omega^4}\left( 1 + \frac{1}{6}(\phi+\phi^\dagger)^2 
		+ \frac{1}{3}\left(b-\frac{1}{2}\right)\left( \phi^2+\phi^{\dagger 2} +(4b+2)|\phi|^2 \right)   \right),\\
	\Omega^2 &=& 1-\frac{1}{6}\left( \phi + \phi^\dagger \right)^2 -\frac{1}{3}\left( b-\frac{1}{2} \right) 
	\left( \phi^2+\phi^{\dagger 2} \right) .
\eea
By minimizing the potential about  $\theta = \theta_{\rm min}$ (\ref{theta_min}) we find that both the K\"ahler metric
and frame function become unity : $\Omega^2 = 1$ and $K_{\phi\bar \phi}=1$.
Thus the resulting scalar potential is simply given by $V = \lambda^2 |\phi|^{2m}$ and chaotic inflation occurs.
This case was noted in Ref.~\cite{Kallosh:2010ug},
which also considered inflation models with a more generic superpotential.
As the value of $b$ increases, the shift symmetry $\phi^2 \to \phi^2 + i\alpha$ appears
and the theory approaches to that studied in Sec.~\ref{case2}.
In particular, for $m=2$, the potential becomes flat and the Higgs-type inflation occurs for sufficiently large field value.
If $b -1/2 \lesssim 3/(4mN)$, the last $N$ e-folds is in the chaotic inflation regime.
Remember that, for a sufficiently large $b$,   the potential exhibits a runaway behavior for $m<2$ and  the potential becomes too steep
for the inflation to occur at large field value for $m>2$.

Next, let us consider the case (ii) ($n>2$).
In this case, the K\"ahler metric and frame function are given by
\bea
	K_{\phi \bar \phi} &=& \frac{1}{\Omega^4}\left( 1- \frac{1}{3}a(\ell-1)^2|\phi |^{2\ell}+a\ell^2|\phi|^{2\ell-2} 
	+\frac{1}{3}b(2\ell-1)(\phi^{2\ell}+\phi^{\dagger 2\ell})  \right. \nonumber \\
	&&\left. ~~~~~+\frac{4}{3}b^2\ell^2|\phi|^{4\ell-2}
	+\frac{1}{3}ab\ell^2|\phi|^{2\ell-2}(\phi^{2\ell}+\phi^{\dagger 2\ell})
	\right), \\
	\Omega^2 &=& 1-\frac{1}{3}\left( |\phi|^2 + a|\phi|^{2\ell}+b(\phi^{2\ell}+\phi^{\dagger 2\ell}) \right).
\eea
In the limit $b\gg |a|$, this approaches to the case of Sec.~\ref{case2}
and the theory has a shift symmetry $\phi^{2\ell} \to \phi^{2\ell} + i\alpha$.
There is another interesting limit $b=\pm 2a$,
where this has an accidental shift symmetry, noting that the frame function is rewritten as
\beq
	\Omega^2 = 1-\frac{1}{3}\left( |\phi|^2 
	+\frac{2b+a}{4}(\phi^{\ell}+\phi^{\dagger \ell})^2
	+\frac{2b-a}{4}(\phi^{\ell}-\phi^{\dagger \ell})^2\right).
\eeq
Thus there appears an approximate shift symmetry, $\phi^\ell \to \phi^\ell + i\alpha$ for $b=a/2$,
and $\phi^\ell \to \phi^\ell + \alpha$ for $b=-a/2$.
The latter case leads to a negative kinetic term for a large field value.
Therefore, we consider the case $b \simeq a/2$ in the following.
Writing $\phi$ as $\varphi e^{i\theta}$, the frame function is given by
\beq
	\Omega^2 = 1-\frac{1}{3}\varphi^2 -\frac{1}{3}\varphi^n\left(a + 2b\cos n\theta\right).
\eeq
The scalar potential is minimized at $\theta = \theta_{\rm min}$ given in (\ref{theta_min}) independently of the sign of $a$
as long as the region connected to the origin without singularities are concerned. 
Then the frame function and K\"ahler metric, along the direction of $\theta = \theta_{\rm min}$, are given by
\bea
	K_{\phi \bar \phi} &=& \frac{1}{\Omega^4}\left( 1+ a\ell^2 \varphi^{2\ell-2}
	 -\frac{1}{3}\left( a\ell^2 + (2b-a)(2\ell-1) \right)\varphi^{2\ell} + \frac{2}{3}b\ell^2(2b-a)\varphi^{4\ell-2}
	  \right), \\
	\Omega^2 &=& 1-\frac{1}{3}\varphi^2 +\frac{1}{3}(2b-a)\varphi^n.
\eea

First, setting $a=2b$, we find that the frame function is simply
reduced to $\Omega^2 = 1-\frac{1}{3}\varphi^2$.  Although the
potential diverges at $\varphi = \sqrt{3}$, a sufficient amount of
inflation still takes place for $\varphi < \sqrt{3}$ as is shown in
the following.  The kinetic term in the Lagrangian takes the following
form,
\beq
	\mathcal L_K = (1+2b \ell^2 \varphi ^{2\ell -2}) (\partial \varphi)^2,
\eeq
in the limit $\varphi \ll \sqrt{3}$.  The canonically normalized field
at large field value is given by
\beq
	\hat \varphi = 2\sqrt{b} \varphi^{\ell}~~~{\rm for}~~~2b\ell^2 \varphi^{2\ell-2} > 1.
\eeq
In the opposite limit $2b\ell^2 \varphi^{2\ell-2} < 1$, the
canonically normalized field is $\tilde \varphi = \sqrt{2}\varphi$.
Thus the scalar potential changes its form as
\beq
	V = \lambda^2 \left(\frac{1}{2\sqrt{b}}\right)^{2m/\ell} \hat\varphi^{2m/\ell}
	~~~{\rm for}~~~2b\ell^2 \varphi^{2\ell-2} > 1,    \label{pot_large}
\eeq
and
\beq
	V = \frac{\lambda^2}{2^m} \tilde\varphi^{2m}  ~~~{\rm for}~~~2b\ell^2 \varphi^{2\ell-2} < 1.   \label{pot_small}
\eeq
This is nothing but the RK inflation model found in
Refs.~\cite{Takahashi:2010ky,Nakayama:2010kt,Nakayama:2010sk}.  One of
the features of the RK inflation is that the power of the potential
becomes smaller at a large field value. In the present case, the power
of the potential during inflation is $2m/\ell$.  In particular, a
fractional power is possible in the RK inflation.  Inflation ends at
$\hat \varphi \sim 1$ and the field value corresponding to the
e-folding number $N$ is $\hat \varphi_N = \sqrt{4mN/\ell}$.\footnote{
  Inflation ends at the large field regime (\ref{pot_large}) if $b >
  2^{\ell-2}\ell^{-2\ell}$.  Otherwise, the last stage of the
  inflation may be described in the small field regime
  (\ref{pot_small}).  In this case we expect a running of the scalar
  spectral index at the scale corresponding to the transition from the
  large to small field regime.  } The corresponding field value of
$\varphi$ is given by $\varphi_N \sim (mN/b\ell)^{1/(2\ell)}$.  This
must satisfy the constraint $\varphi_N < \sqrt{3}$ for the above
analysis to be valid.  This translates into the bound on $b$ as
\beq
	b > \frac{mN}{3^\ell \ell}.
\eeq
The inflaton dynamics and the corresponding thermal history after
inflation in the RK inflation model have been studied in detail in
\cite{Nakayama:2010kt} and not repeated here.  We only show the
spectral index and the tensor to scalar ratio,
\beq
	n_s \simeq 1-\left(1+\frac{m}{\ell}\right)\frac{1}{N},~~~
	r  \simeq \frac{8m}{\ell}\frac{1}{N}.
\eeq

In the above analysis we assumed $2b=a$.  Let us see how the dynamics
is affected if this equality is violated.  One can show that if the
following condition is satisfied,
\beq
	b - \frac{a}{2} > b_c \equiv \frac{(n-2)^\frac{n-2}{2}}{3^\frac{n-2}{2} n^{\frac{n}{2}}},
\eeq
the scalar potential does not diverge along the direction $\theta = \theta_{\rm min}$.
There is a local maximum of the potential along $\theta = \theta_{\rm min}$,
which may be an obstacle to the inflation.
The condition that the local maximum disappears is written as
\beq
	b-\frac{a}{2} > b_c' = \frac{(m-2)^\frac{n}{2}}{m^\frac{n-2}{2} (m-n)} b_c.
\eeq
One can show that $b_c' \geq b_c $ for $2 \leq n < m$.
The dynamics of the RK inflation is not much affected as long as
$|b-a/2| \lesssim \varphi_N^{-2\ell} \sim b\ell/(mN)$.  Otherwise,
if $b$ is sufficiently large, higher order terms in the frame function and K\"ahler metric becomes
important, and the theory approaches to that studied in
Sec.~\ref{case2}.

To summarize, the RK inflation is realized when $|b-a/2| \lesssim
b\ell/(mN)$, and the Higgs-type or power-law inflation is realized
when $b-a/2 \gg b_c'$ for certain choices
of $m$ and $n$.  In the case of $b-a/2 \gg b_c'$ but $|b-a/2| \lesssim
b\ell/(mN)$, the power-law inflation is followed by the RK inflation
for the last $N$ e-foldings.

\section{Discussion and Conclusions}
We have studied the inflation models in Jordan frame supergravity with
a logarithmic K\"ahler potential, and found that the three types of
inflation are possible: the Higgs-type, the power-law and the RK
inflation, depending on the relative importance of the holomorphic and
non-holomorphic functions in the frame function. More precisely
speaking, when the holomorphic function is important, the potential
exhibits runaway behavior for $m<n$ and $m=n\geq 3$, and the inflation
does not occur. The Higgs-type inflation occurs if $m=n=2$, and the
power-law inflation takes place if $0<(m-n)/n \ll 1$. We have pointed
out that, in this case, there is a shift symmetry on a holomorphic
function, which is basically equivalent to (\ref{sym}) considered in
the RK inflation. Although the dynamics is not same because of the
logarithmic form of the K\"ahler potential, it is remarkable that the
inflationary path is same as that considered in
Ref.~\cite{Nakayama:2010kt}. On the other hand, if the non-holomorphic
function and the holomorphic one are comparable to each other, there
appears another shift symmetry.  In the case of $n=2$, this leads to a
usual chaotic inflation, while the RK inflation
is realized for $n=2\ell > 2$. Interestingly, a fractional power
potential is possible for the RK inflation due to the running kinetic
term. In particular, we have shown that the same dynamics considered in
Refs.~\cite{Takahashi:2010ky,Nakayama:2010kt,Nakayama:2010sk} is
realized with the logarithmic K\"ahler potential.

\begin{figure}[t]
\includegraphics[scale=0.5]{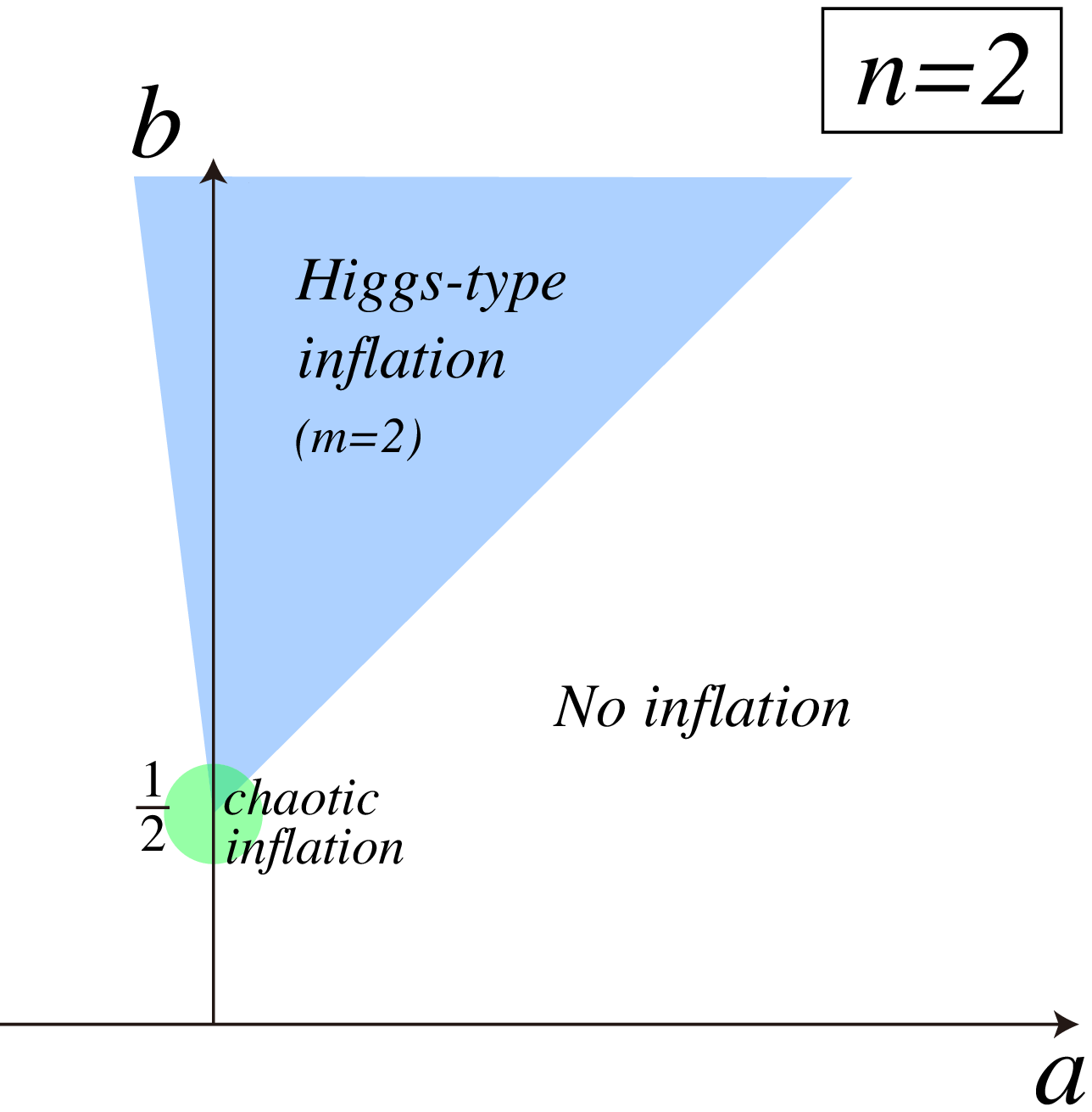}
\includegraphics[scale=0.5]{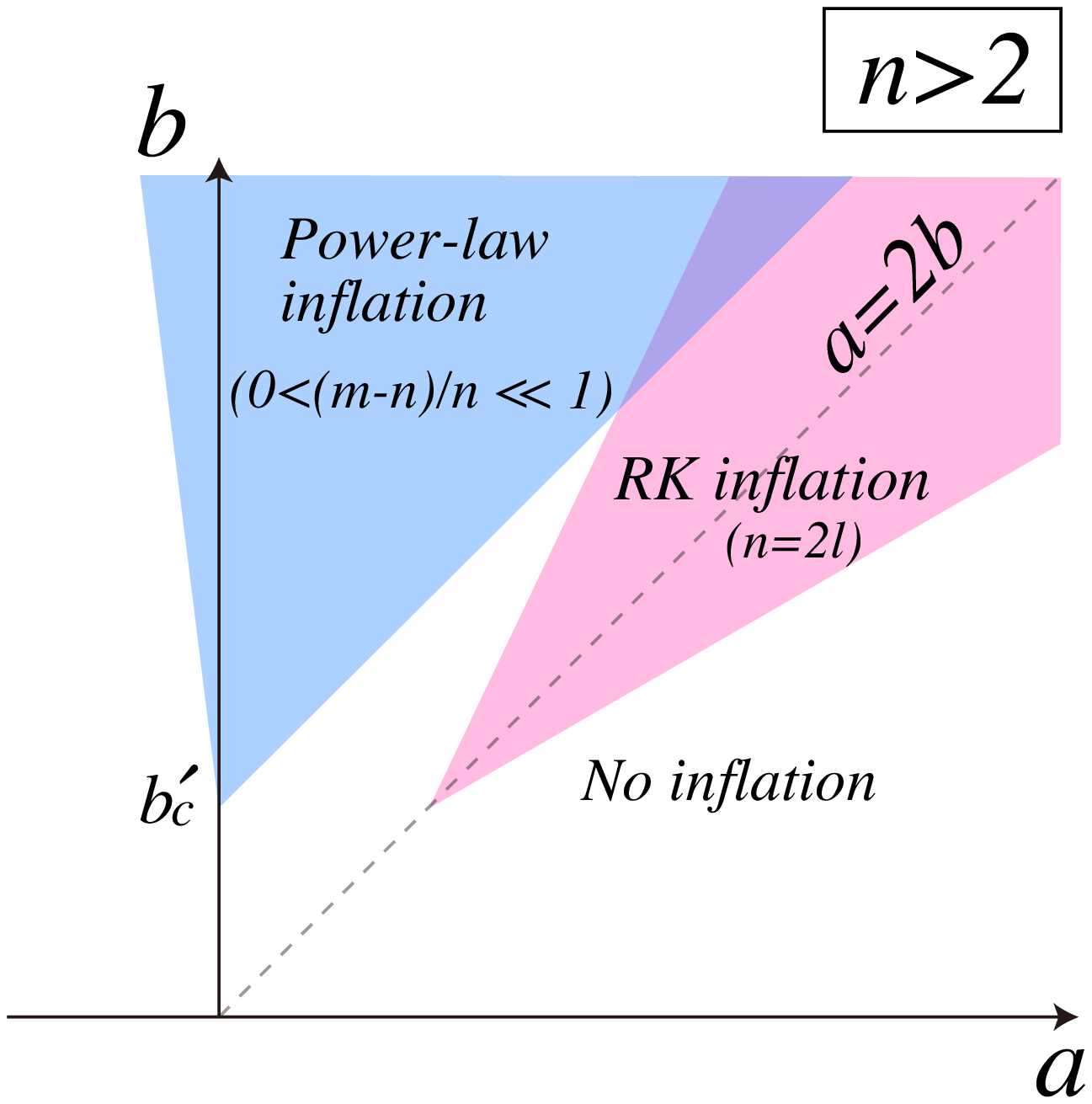}
\caption{The type of inflation models realized in supergravity with
  the logarithmic K\"ahler potential. Here $a$ and $b$ denote the
  coefficients of the non-holomorphic and holomorphic functions,
  respectively. See the text for details.
  Scales of both axes are arbitrary, and so,
  the area of each  region is not necessarily  proportional to 
  the likelihood.
    }
\label{fig:a-b}
\end{figure}

The relation of the inflation models is schematically shown in
Fig.~\ref{fig:a-b}.  In the left panel corresponding to the case of
$n=2$, the Higgs-type inflation is possible for sufficiently large $b$
(blue triangle), while the chaotic inflation with the potential $\propto \varphi^{2m}$ occurs for $b=1/2$ and
$a\approx 0$ (green circle) because there appears a shift
symmetry. We note that, if $b$ is sufficiently
large, the Higgs-type inflation is possible for any $\ell$ and $a$. In
the right panel corresponding to $n>2$, the power-law inflation occurs
for large $b$ (blue triangle), while there appears an approximate
shift symmetry along $b=2a$, leading to the RK inflation with with potential $\propto \varphi^{2m/\ell}$ if $n=2\ell$.
Interestingly, in the overlapping region, the power-law inflation
takes place, subsequently followed by the RK inflation at smaller
field values, as in
Refs.~\cite{Takahashi:2010ky,Nakayama:2010kt}. Note that, if $n \ne 2
\ell$ and $n > 2$, only the power-law inflation is possible for
sufficiently large $b$.

Lastly we briefly mention the implication for the Higgs inflation in
supergravity.  In NMSSM, there is an interaction of the Higgs fields
and an additional singlet $S$,
\beq
W\;=\; \lambda S H_u H_d,
\eeq
which is same as (\ref{Xphi}) with $m=2$, noting that $H_uH_d$ can be
described as $\phi^2$ along the D-flat direction. The interaction
generates a quartic coupling about the origin. There are two issues in
using the Higgs fields as the inflaton, if we extrapolate the quartic
potential up to a large field value. First, the chaotic inflation with
a quartic potential is excluded by the WMAP observation.  Second, the
coupling needed for obtaining the correct magnitude of the density
perturbation is very small, $\lambda \sim 10^{-6}$. As to the first
issue, we need to somehow make the potential flatter at a large field
value in order to realize a Higgs inflation that is consistent with
observation. As far as we know, there are two ways; one is to
introduce a non-minimal coupling to gravity, and the other is to
consider a running kinetic term~\cite{Nakayama:2010sk}. As is well
known, the former leads to a potential given by a constant plus an
exponentially suppressed tilt, while the latter enables e.g. quadratic
or even fractional-power potentials. These two possibilities predict
different tensor-to-scalar ratio $r$, and the RK Higgs inflation tends
to predict a larger $r$ within the reach of the Planck
satellite~\cite{:2006uk}.  Concerning the second issue on the small
coupling, one needs either a large non-minimal coupling to gravity or
a small $\lambda$ in the former case~\cite{Lee:2010hj}. On the other
hand, $\lambda$ can be as large as $O(0.1)$ without generating a large $\mu$ term in the 
RK Higgs inflation.  
This is because the kinetic term after inflation is different from
that during inflation. Applying our analysis in this letter to the
case of the Higgs inflation, we can see that the two possibilities,
the non-minimally coupled Higgs inflation and the RK Higgs inflation,
are related to each other. Their difference arises from the relative
size of the holomorphic and non-holomorphic functions.
Note that the power-law inflation does not take place because $1<(m-n)/n \ll 1$ cannot be satisfied for $m=2$.

%
\begin{acknowledgements}
  
  This work was supported by the Grant-in-Aid for Scientific Research
  on Innovative Areas (No. 21111006) [KN and FT] and Scientific
  Research (A) (No. 22244030) [FT], and JSPS Grant-in-Aid for Young
  Scientists (B) (No. 21740160) [FT].  This work was supported by
  World Premier International Center Initiative (WPI Program), MEXT,
  Japan.

\end{acknowledgements}



\end{document}